\documentclass[english,prd,floatfix,superscriptaddress,nofootinbib]{revtex4-2}
\usepackage[T1]{fontenc}
\usepackage[utf8]{inputenc}
\usepackage{color}
\definecolor{note_fontcolor}{rgb}{0.800781, 0.800781, 0.800781}
\usepackage{babel}
\usepackage{mathtools}
\usepackage{amsmath}
\usepackage{amssymb}
\usepackage{graphicx}
\usepackage{rotfloat}
\usepackage{booktabs}
\PassOptionsToPackage{hyphens}{url}\usepackage[unicode=true,pdfusetitle,
bookmarks=true,bookmarksnumbered=false,bookmarksopen=false,
breaklinks=false,pdfborder={0 0 0},pdfborderstyle={},backref=false,colorlinks=false]{hyperref}

\makeatletter

\usepackage{colortbl}
\usepackage{upgreek}
\usepackage{url}

\usepackage{tensind}
\tensordelimiter{?}
\usepackage{pgfplots}
\usetikzlibrary{pgfplots.groupplots}

\DeclareSymbolFont{vectors}{OML}{cmm}{b}{it}
\DeclareSymbolFont{tensors}{OT1}{cmss}{bx}{it}

\DeclareSymbolFontAlphabet{\mathvec}{vectors}
\DeclareSymbolFontAlphabet{\mathtens}{tensors}
\DeclareUnicodeCharacter{03B2}{\ensuremath{\upbeta}}

\usepackage[capitalise]{cleveref}

\makeatother

\begin{document}
\global\long\def\tudu#1#2#3#4{{?{#1}^{#2}{}_{#3}{}^{#4}?}}%

\global\long\def\tud#1#2#3{{?{#1}^{#2}{}_{#3}?}}%

\global\long\def\tudud#1#2#3#4#5{{?{#1}^{#2}{}_{#3}{}^{#4}{}_{#5}?}}%

\global\long\def\tdu#1#2#3{\tensor{#1}{_{#2}^{#3}}}%

\global\long\def\dd#1#2{\frac{\mathrm{d}#1}{\mathrm{d}#2}}%

\global\long\def\pd#1#2{\frac{\partial#1}{\partial#2}}%

\global\long\def\tens#1{\mathtens{#1}}%

\global\long\def\threevec#1{\mathvec{#1}}%

\global\long\def\d{\mathrm{d}}%

\global\long\def\e{\mathrm{e}}%

\global\long\def\eps{\varepsilon}%

\global\long\def\i{\mathrm{i}}%

\global\long\def\ext{\tilde{\mathrm{d}}}%

\title{\bf Black holes in scalar multipolar Universes}
\author{Carlos A. R. Herdeiro}
\email{herdeiro@ua.pt}
\affiliation{Departamento de Matem\'atica da Universidade de Aveiro and Centre for Research and Development  in Mathematics and Applications (CIDMA), Campus de Santiago, 3810-193 Aveiro, Portugal.}

\begin{abstract}
Recently, a scalar counterpart of the Schwarzschild-Melvin Universe was reported~\cite{Cardoso:2024yrb}. We show this solution is a special case of a Schwarzschild black hole/mass in a scalar multipolar Universe, that can be constructed \textit{algebraically combining} known vacuum solutions. This builds on the generalized Weyl construction for scalar-vacuum, that admits two, fully decoupled, independent harmonic functions: one for the gravitational sector and another for the scalar sector. Harmonic solutions with growing multipoles lead to naked singularities (at spatial infinity), including for the scalar counterpart of the Schwarzschild-Melvin; harmonic solutions with decaying (scalar) multipoles are asymptotically flat, but have a singular horizon, in accordance with no-scalar-hair theorems. The scalar-vacuum model can be mapped to a five dimensional pure gravity construction via Kaluza-Klein oxidization and, in this way, to the corresponding $D=5$ generalized Weyl construction.
\end{abstract}

\maketitle

\section{Introduction}
The Melvin magnetic Universe is an everywhere smooth, non-asymptotically flat, solution to the Einstein-Maxwell theory~\cite{Bonnor_1954,Melvin:1963qx}. It lends itself to many applications and generalizations ($e.g.$~\cite{Tseytlin:1994ei,Costa:2000nw,Costa:2001ifa,Gibbons:2001sx,Ortaggio:2003ri,Radu:2003av,Astorino:2012zm,Bambi:2015sla,Zofka:2019yfa,Bronnikov:2020ikh,Kastor:2020wsm,Sabra:2023fps}), including accommodating a Schwarzschild~\cite{Ernst:1976mzr} (or Reissner-Nordstr\"om~\cite{Gibbons:2013yq} or Kerr~\cite{Ernst:1976bsr}) black hole (BH).

In a recent paper~\cite{Cardoso:2024yrb}, an  exact solution to the Einstein-scalar field equations was reported, described as a \textit{scalar counterpart} of the Schwarzschild-Melvin solution. The solution was reportedly found by inspection. It may therefore be pedagogical to explain how this solution is a particular case of a more general family of Weyl solutions, constructed by choosing appropriate multipolar fields in the associated Weyl construction, that yields two harmonic functions in an auxiliary space, one associated with the gravitational sector and another with the scalar sector. In each sector, on the one hand, the non-linearities of the field equations perdure in the way the harmonic function sources the remaining metric function. On the other hand, the way the two sectors source the remaining metric function is \textit{linear}, and admits a superposition principle.

As in electrostatics, the harmonic equation can be solved by a multipolar expansion. These multipoles can be growing or decaying and any of these can be superimposed with a BH in a systematic way, due to the linearity of the two associated Weyl harmonic problems. The solution in~\cite{Cardoso:2024yrb} is obtained by choosing the growing dipole (in the auxiliary space) for the scalar sector, superimposed with the Schwarzschild BH, obtained in a standard way as the Newtonian potential of a rod (in the auxiliary space), for the gravitational sector. Unfortunately, as for all solutions with growing multipoles, both in the gravitational and scalar sectors, it has a curvature singularity at spatial infinity. Moreover, at least for the solution in~\cite{Cardoso:2024yrb}, this singularity is in causal contact with the bulk, being therefore naked. This creates an important distinction with the everywhere regular Melvin Universe. 

Still, both the scalar counterpart of the Melvin Universe and its multipolar generalizations (with growing multipoles) allow non-trivial scalar multipoles that appear smooth in the vicinity of the horizon, something forbidden for asymptotically flat solutions in this theory, as established by well-known no scalar no-hair theorems~\cite{Chase:1970omy,Bekenstein:1972ny,Herdeiro:2015waa}. Therefore, these solutions may be interesting as a proxy to explore scalar-BH interactions in that spacetime region, possibly considering a large distance cut-off to remove the singularity at infinity.

\section{The Weyl construction for vacuum and scalar-vacuum}
General Relativity is a non-linear field theory. In particular circumstances, however, it admits a partial, or even full, linearization. The pioneering example is the Weyl construction~\cite{Weyl1917} for vacuum axi-symmetric solutions. Using the \textit{static and axisymmetric} ansatz
\begin{eqnarray}
    ds^2&=&-e^{2U(\rho,z)}dt^2+e^{-2U(\rho,z)}\left[e^{2\gamma(\rho,z)}(d\rho^2+dz^2)+\rho^2d\varphi^2\right] \ ,
    \label{ansatzweyl}
\end{eqnarray}
in Weyl canonical coordinates $(t,\rho,z,\varphi)$, the vacuum field equations $R_{\mu\nu}=0$ yield the Laplace equation
\begin{equation}
    \Delta_{\mathbb{E}^3}U(\rho,z)=0 \ ,
    \label{Weylharmonic}
\end{equation}
on the auxiliary Euclidean 3-space $ds^2_{\mathbb{E}^3}=d\rho^2+\rho^2d\varphi^2+dz^2$ and the two line integrals
\begin{eqnarray}
\gamma_{,\rho} =\rho  \left[ (U_{,\rho})^2-(U_{,z})^2\right] 
\ , \qquad 
\gamma _{,z}=2\rho U_{,z} U_{,\rho} \ .
\label{Weylline}
\end{eqnarray}
Then, the methodology to solve the Weyl problem is to choose a harmonic function $U(\rho,z)$, which can be thought of as a Newtonian gravitational potential of sources along the $z$-axis, and obtain the corresponding $\gamma=\gamma[U]$ by quadrature, solving eqs.~\eqref{Weylline}.

The linear harmonic structure observed in eq.~\eqref{Weylharmonic} allows many interesting constructions. For instance one can superimpose two~\cite{Bach1922} or many~\cite{Israel:1964mgs} Schwarzschild BHs; one can also have solutions corresponding to pure multipoles \textit{in the auxiliary space}, which can be growing or decaying. As an example, the decaying monopole is a naked singularity called Chazy-Curzon particle, with a non-trivial multipolar structure in spacetime~\cite{Chazy1924,Curzon25}. 
Furthermore, one can also superimpose BHs with such gravitational multipoles~\cite{Astorino:2021dju,Vigano:2022hrg}. An algorithmic procedure, known as the inverse scattering construction~\cite{Belinsky:1979mh,Harmark:2004rm}, allows furthermore such solutions to be used as seeds to construct rotating counterparts, $i.e.$ stationary (but not static) and axi-symmetric solutions. 

One of the simplest generalizations of this Weyl construction is to add a massless, free scalar field $\Phi$, minimally coupled to Einstein's gravity (aka \textit{scalar-vacuum}):\footnote{We use units with $G=c=1$. The scalar field herein is scaled by a factor of $\sqrt{4\pi}$ with respect to the one in~\cite{Cardoso:2024yrb}.}
\begin{equation}
 \mathcal{S}= \frac{1}{16\pi}\int d^{4}x \sqrt{-g}\left(R-2\partial_\mu{\Phi} \partial^\mu {\Phi}\right) \ .
    \label{actionEinsca}
\end{equation}
Then, under the same metric ansatz~\eqref{ansatzweyl}, and the scalar field ansatz
\begin{equation}
    \Phi=\Phi(\rho,z) \ ,
    \label{sansatz}
\end{equation}
the field equations become\footnote{These can be obtained, $e.g.$, as a special case of the model considered in~\cite{Herdeiro:2023mpt}.}
\begin{equation}
    \Delta_{\mathbb{E}^3}U(\rho,z)=0 \ , \qquad  \Delta_{\mathbb{E}^3}\Phi(\rho,z)=0 \ ,
\end{equation}
 and the two line integrals to determine $\gamma$ now a function of $U$ and $\Phi$, denoted as $\gamma=\gamma[U;\Phi]$:
\begin{eqnarray}
\gamma_{,\rho}=\rho  \left[(U_{,\rho})^2-(U_{,z})^2+(\Phi_{,\rho})^2-(\Phi_{,z})^2\right]
\ , \qquad 
\gamma _{,z}=2\rho\left[ \Phi_{,\rho} \Phi_{,z}+ U_{,\rho} U_{,z}\right] \ .
\end{eqnarray}
Two key observations are the following. Firstly, there are \textit{two} independent harmonic functions in the auxiliary space, one for the gravitational sector, $U(\rho,z)$, and one for the scalar sector, $\Phi(\rho,z)$. Both of them source \textit{democratically}  the line integrals that determine the remaining metric function $\gamma(\rho,z)$. In each sector this sourcing is \textit{non-linear}, meaning that if $U_1$, $U_2$ ($\Phi_1$, $\Phi_2$) are harmonic solutions of $\Delta_{\mathbb{E}^3}U(\rho,z)=0$ ($\Delta_{\mathbb{E}^3}\Phi(\rho,z)=0$), then 
\begin{eqnarray}
    \gamma[U_1+U_2;0]\neq \gamma[U_1;0]+\gamma[U_2;0] \ ,  \\ \qquad  \gamma[0;\Phi_1+\Phi_2]\neq \gamma[0;\Phi_1]+\gamma[0;\Phi_2] \ . 
\end{eqnarray}
Thus, within each sector, there is only a partial linearization of the field equations. Secondly, the gravitational and scalar sector source \textit{independently} $\gamma$:
\begin{equation}
    \gamma[U;\Phi]= \gamma[U;0]+\gamma[0;\Phi] \ . 
\end{equation}
Thus, the scalar and gravitational sectors \textit{decouple} and there is a full linearization of the Einstein equations, and a corresponding inter-sectorial superposition principle.

This construction unveils two solution generating techniques, \textit{simply combining} or \textit{redistributing} known vacuum solutions of the form~\eqref{ansatzweyl}, determined by a choice of harmonic functions $U$, here denoted $g=g[U,\gamma[U] ]$, yielding solutions of~\eqref{actionEinsca} of the form $\{g;\Phi\}=\{g[U,\gamma[U]+\gamma[\Phi]];\Phi\}$:
\begin{description}
      \item[i)] Selecting \textit{two} vacuum solutions $g=g[U_1,\gamma_1[U_1]]$ and $g=g[U_2,\gamma_2[U_2]]$, these can be combined into a scalar-vacuum solution  $\{g[U_1,\gamma[U_1]+\gamma[U_2]];\Phi=U_2\}$. This can be used straightforwardly to generate the solution in~\cite{Cardoso:2024yrb} and generalizations thereof from known vacuum solutions, as shown below;
     \item[ii)] Any vacuum solution $g=g[U,\gamma[U] ]$ is mapped to a scalar-vacuum solution of~\eqref{actionEinsca} of the form $\{g[0,\gamma[U]];\Phi=U\}$. This is a special case of the Buchdahl reciprocal solutions~\cite{Buchdahl:1959nk}, with $\beta=0$ in the notation therein.
\end{description}
We shall now use procedure ${\bf i)}$ to obtain scalar-vacuum solutions from known vacuum solutions.

\section{Scalar-vacuum solutions \\ from the Schwarzschild BH and gravitational multipoles}

It is well known that the Schwarzschild BH of mass $M$ emerges in the Weyl construction taking $U (\rho,z)=U_{\rm rod}(\rho,z)$, the Newtonian potential of a finite, constant (line) mass density, zero thickness rod, along the $z$-axis, with length $2M$. Centering it at $z=0$, the potential reads~\cite{Stephani:2003tm}
\begin{equation}
    U_{\rm rod}(\rho,z)=\frac{1}{2}\ln\left[\frac{r_++r_--2M}{r_++r_-+2M}\right]  , \qquad r^2_\pm\equiv \rho^2+(z\pm M)^2 \ . 
\end{equation}
The line integrals~\eqref{Weylline} then yield
\begin{equation}
    \gamma_{\rm rod}(\rho,z)=\frac{1}{2}\ln\left[\frac{(r_++r_--2M)(r_++r_-+2M)}{4r_+ r_-}\right]  \ . 
\end{equation}
Applying to~\eqref{ansatzweyl} with $U= U_{\rm rod}$ and $\gamma=\gamma_{\rm rod}$ the coordinate transformation $(\rho,z)\rightarrow (r,\theta)$ given by
\begin{equation}
  z(r,\theta)=(r-M)\cos\theta \ , \qquad \rho(r,\theta)=\sqrt{r^2-2Mr}\sin\theta \ ,  
  \label{ctweyl}
\end{equation}
one recovers the Schwarzschild solution in the canonical form
\begin{eqnarray}
    ds^2=-f(r)dt^2+\frac{dr^2}{f(r)}+r^2d\theta^2+r^2\sin^2\theta d\varphi^2 \ , \qquad    f(r)=1-\frac{2M}{r} \ .
\end{eqnarray}

Another vacuum solution, corresponds to taking the standard (axi-symmetric) multipolar expansion on ${\mathbb{E}^3}$, which in standard cylindrical coordinates $(\rho,z,\varphi)$ on ${\mathbb{E}^3}$  reads~\cite{Stephani:2003tm}
\begin{equation}
    U_{\rm multi}(\rho,z)
    =
    \frac{a_0}{R}
    +\sum_{\ell=1}^\infty \left(
    \frac{a_\ell}{R^{\ell+1}}
    +b_\ell R^\ell\right)P_\ell \left[\frac{z}{R}
    \right] \ ,
    \label{Umulti}
\end{equation}
where $R=\sqrt{\rho^2+z^2}$ and $P_\ell[x]$ are the  Legendre polynomials with argument $x$. The line integrals~\eqref{Weylline} then yield~\cite{Vigano:2022hrg} (omitting the argument of the Legendre polynomials):
\begin{equation}
    \gamma_{\rm multi}(\rho,z)=\sum_{\ell,m=0}^\infty\left[ 
    \frac{(\ell+1)(m+1)a_\ell a_m}{(\ell+m+2)R^{\ell+m+2}}(P_{\ell+1}P_{m+1}-P_\ell P_m)\right]
    +\sum_{\ell,m=1}^\infty\left[ 
    \frac{\ell m b_\ell b_m R^{\ell+m}}{\ell+m}(P_{\ell}P_{m}-P_{\ell-1} P_{m-1})
    \right] \ . 
    \label{gammageneral}
\end{equation}

Consequently, a general family of scalar-vacuum solutions, determined by the parameters $(M,a_\ell,b_\ell$) and interpreted as a Schwarzschild mass/BH immersed in scalar multipolar fields is obtained - by the procedure {\bf i)} above - taking
\begin{equation}
    U(\rho,z)= U_{\rm rod}(\rho,z) \ , \qquad  \Phi(\rho,z)= U_{\rm multi}(\rho,z) \ , \qquad \gamma(\rho,z)= \gamma_{\rm rod}(\rho,z)+ \gamma_{\rm multi}(\rho,z) \ ,
\end{equation}
all the information being explicitly written above. With this construction, the coordinate transformation~\eqref{ctweyl} brings the solution to the form
\begin{equation}
    ds^2=-f(r)dt^2+e^{2\gamma_{\rm multi}(r,\theta)}\left[\frac{dr^2}{f(r)}+r^2d\theta^2\right]+r^2\sin^2\theta d\varphi^2 \ , \qquad    \Phi(r,\theta)= U_{\rm multi}(r,\theta)  \ ,
    \label{BHmulti}
\end{equation}
where
\begin{equation}
    R(r,\theta)=r\sqrt{f(r)+\frac{M^2\cos^2\theta}{r^2}} \ .
    \label{Rfunction}
\end{equation}
The scalar-vacuum solution~\eqref{BHmulti} describes a Schwarzschild BH for $a_\ell=0=b_\ell$, $\ell\in \mathbb{N}_0$. For $M=0$ it describes a scalar environment, with decaying or growing multipoles (in the auxiliary) space. In the following we analyse some special cases.

\section{A Schwarzschild metric in a single monopole scalar environments}
We consider that there is only one decaying multipole (say $a_\ell$) or one growing multipole (say $b_\ell$). Then we have for the former
\begin{equation}
  U_{\rm multi}^{[a_\ell]}(r,\theta)= 
  \frac{a_\ell}{[R(r,\theta)]^{\ell+1}}P_\ell\left[\frac{z(r,\theta)}{R(r,\theta)}\right] \ , 
  \qquad 
   \gamma_{\rm multi}^{[a_\ell]}(r,\theta)=
    \frac{(\ell+1)a_\ell^2}{2[R(r,\theta)]^{2\ell+2}}\left\{\left(P_{\ell+1}\left[\frac{z(r,\theta)}{R(r,\theta)}\right]\right)^2-\left(P_\ell\left[\frac{z(r,\theta)}{R(r,\theta)}\right]\right)^2\right\} \ ,
    \label{decay}
\end{equation}
whereas for the latter we have
\begin{equation}
  U_{\rm multi}^{[b_\ell]}(r,\theta)
  = {b_\ell}{[R(r,\theta)]^{\ell}}P_\ell\left[\frac{z(r,\theta)}{R(r,\theta)}\right] \ , 
  \qquad 
   \gamma_{\rm multi}^{[b_\ell]}(r,\theta)=
    \frac{\ell b_\ell^2 [R(r,\theta)]^{2\ell}}{2}\left\{\left(P_{\ell}\left[\frac{z(r,\theta)}{R(r,\theta)}\right]\right)^2-\left(P_{\ell-1}\left[\frac{z(r,\theta)}{R(r,\theta)}\right]\right)^2\right\} \ . 
     \label{growing}
\end{equation}
Let us now choose specific values of $\ell$.

\subsection{$a_0$: (decaying) monopole}

Consider only the (decaying) monopole in~\eqref{decay}, $a_0\neq 0$. Then
\begin{equation}
  U_{\rm multi}^{[a_0]}(r,\theta)= 
  \frac{a_0}{\sqrt{r^2f(r)+M^2\cos^2\theta}} \ , 
  \qquad 
   \gamma_{\rm multi}^{[a_0]}(r,\theta)=
    -\frac{a_0^2 f(r) r^2 \sin ^2\theta }{2 \left[r^2f(r)+M^2\cos^2\theta\right]^2} \ ,
    \label{a0}
\end{equation}
so that the solution reads
\begin{equation}
    ds^2=-f(r)dt^2+e^{-\frac{a_0^2 f(r) r^2 \sin ^2\theta }{ \left[r^2f(r)+M^2\cos^2\theta\right]^2}}\left[\frac{dr^2}{f(r)}+r^2d\theta^2\right]+r^2\sin^2\theta d\varphi^2 \ , \qquad    \Phi(r,\theta)= \frac{a_0}{\sqrt{r^2f(r)+M^2\cos^2\theta}}  \ .
    \label{sola0}
\end{equation}
This solution represents a Schwarzschild mass inside a scalar version of the Chazy-Curzon particle. Observe that even though in the auxiliary space we have chosen a single monopole, in spacetime there is a non-trivial multipolar dependence (for the scalar field). Moreover, the would be horizon is now singular, as can be seen from the Ricci scalar
\begin{equation}
    R=\frac{ 2a_0^2 \left[r^2f(r)+M^2\sin^2\theta \right] }{r^2 \left[r^2f(r)+M^2\cos^2\theta\right]^2} \exp \left[\frac{a_0^2 f(r) r^2 \sin ^2\theta}{\left(r^2f(r)+M^2\cos^2\theta\right)^2}\right]\ ,
\end{equation}
which blows up at $r=2M$ and $\theta=\pi/2$.

\subsection{$b_1$: (growing) dipole}
Consider only the growing dipole in~\eqref{growing}, $b_1\neq 0$. Then
\begin{equation}
  U_{\rm multi}^{[b_1]}(r,\theta)= 
  b_1(r-M)\cos\theta \ , 
  \qquad 
   \gamma_{\rm multi}^{[b_1]}(r,\theta)=
    -b_1^2 f(r) r^2 \sin ^2\theta \ ,
    \label{b1}
\end{equation}
so that the solution reads
\begin{equation}
    ds^2=-f(r)dt^2+e^{-b_1^2 f(r) r^2 \sin ^2\theta}\left[\frac{dr^2}{f(r)}+r^2d\theta^2\right]+r^2\sin^2\theta d\varphi^2 \ , \qquad    \Phi(r,\theta)=  b_1(r-M)\cos\theta   \ .
    \label{solb1}
\end{equation}
This is the scalar counterpart of the Schwarzschild-Melvin solution, in the terminology of~\cite{Cardoso:2024yrb}. However, unlike the Melvin Universe of Einstein-Maxwell theory, this scalar Universe is \textit{not} everywhere regular. For instance, the Ricci scalar reads
\begin{equation}
 R= \frac{2b_1^2 \left[r^2f(r)+M^2\sin^2\theta\right] e^{b_1^2 f(r)r^2 \sin ^2\theta}}{r^2} \ ,
\end{equation}
which diverges as $r\rightarrow \infty$, except along the axis where $\sin\theta=0$. Thus, $r=\infty$ is a physical singularity. Moreover, it is in causal contact with the whole spacetime. For instance, equatorial photons with energy $E>0$ and angular momentum $j$ obey
\begin{equation}
    \left(\frac{dr}{d\lambda}\right)^2=e^{b_1^2 f(r) r^2}\left[E^2-\frac{j^2}{r^2}f(r)\right] \ .
\end{equation}
Thus, in the empty scalar Universe ($M=0$), for instance, a radial equatorial photon travels from the origin to radial infinity in a finite affine parameter
\begin{equation}
    \Delta \lambda=\frac{1}{E}\int_0^\infty e^{-b_1^2r^2/2}dr=\sqrt{\frac{\pi}{2b_1^2E^2}} \ ,
\end{equation}
making the singularity \textit{naked}.

Still concerning the motion of light rays, one can also see that, unlike the Melvin Universe~\cite{Junior:2021dyw}, the scalar Universe \textit{does not} introduce a light ring. Indeed, light rings are critical points of a certain potential~\cite{Cunha:2020azh}, which reads, for the metric~\eqref{solb1},
\begin{equation}
    H=\sqrt{\frac{-g_{tt}}{g_{\varphi\varphi}}}=\frac{\sqrt{f}}{r\sin\theta} \ .
\end{equation}
This is the same as Schwarzschild. So the only light ring is at $r=3M$. Moreover, due to the breakdown of the spherical symmetry, there is only one equatorial light ring, rather than the Schwarzschild  photon sphere.

\subsection{$a_1$: (decaying) dipole}
Consider now, instead, only the decaying dipole in~\eqref{decay}, $a_1\neq 0$. Then, 
\begin{equation}
  U_{\rm multi}^{[a_1]}(r,\theta)= 
  \frac{a_1(r-M)\cos\theta}{[r^2f(r)+M^2\cos^2\theta]^{3/2}} \ , 
  \qquad 
   \gamma_{\rm multi}^{[a_1]}(r,\theta)=
    -\frac{a_1^2 f(r) r^2 \sin ^2\theta[r^2 f(r) (8 \cos^2 \theta -\sin^2\theta)+8 M^2 \cos ^2\theta] }{4 \left[r^2f(r)+M^2\cos^2\theta\right]^4} \ .
    \label{a1}
\end{equation}
so that the solution reads
\begin{equation}
    ds^2=-f(r)dt^2+e^{2\gamma_{\rm multi}^{[a_1]}(r,\theta)}\left[\frac{dr^2}{f(r)}+r^2d\theta^2\right]+r^2\sin^2\theta d\varphi^2 \ , \qquad    \Phi(r,\theta)=  \frac{a_1(r-M)\cos\theta}{[r^2f(r)+M^2\cos^2\theta]^{3/2}}   \ .
    \label{sola1}
\end{equation}
The Ricci scalar is now cumbersome. At $r=2M$ it reads
\begin{equation}
    R=\frac{2a_1^2 \tan ^2\theta}{M^6\cos^4\theta} \ ,
\end{equation}
so that we see again that this surface is singular along the equator. 

\subsection{$b_2$: (growing) quadrupole}
As a final example we consider only the growing quadrupole in~\eqref{growing}, $b_2\neq 0$. Then
\begin{eqnarray}
  U_{\rm multi}^{[b_2]}(r,\theta)= 
  \frac{b_2}{2}  \left[2 M^2 \cos ^2\theta +r^2f(r) \left(2 \cos
   ^2\theta -\sin ^2\theta \right)\right] \ , \\ 
   \gamma_{\rm multi}^{[b_2]}(r,\theta)=
    -\frac{b_2^2 f(r) r^2 \sin ^2\theta[r^2 f(r) (8 \cos^2 \theta -\sin^2\theta)+8 M^2 \cos ^2\theta] }{4} \ .
    \label{b2}
\end{eqnarray}
so that the solution reads
\begin{equation}
    ds^2=-f(r)dt^2+e^{2\gamma_{\rm multi}^{[b_2]}(r,\theta)}\left[\frac{dr^2}{f(r)}+r^2d\theta^2\right]+r^2\sin^2\theta d\varphi^2 \ , \qquad    \Phi(r,\theta)=  U_{\rm multi}^{[b_2]}(r,\theta)  \ .
    \label{solb2}
\end{equation}
The Ricci scalar is again cumbersome. But one can check it diverges at spatial infinity for some angles. For instance, along the equatorial plane ($\theta=\pi/2$) and the ``cone" $\theta=\pi/4$ it reads
\begin{eqnarray}
    R\left(\theta=\frac{\pi}{2}\right)&=&2 b_2^2 (M-r)^2 f(r) e^{-\frac{b_2^2}{2}  r^4 f(r)^2} \ , \\
    R\left(\theta=\frac{\pi}{4}\right)&=&
    \frac{b_2^2 \left[M^2+2 r^2f(r)\right] \left[4 M^2+5 r^2f(r)\right] e^{\frac{b_2^2}{8}  r^2 f(r) \left[8 M^2+7 r^2f(r)\right]}}{2 r^2}  \ ,
\end{eqnarray}
so that it diverges at spatial infinity along latter, but not the former.
\subsection{Summary}

The following table summarizes these four first multipoles, using the function $R(r,\theta)$, $cf.$ eq.~\eqref{Rfunction} for compactness, where, in particular, the equivalence between the functional form of $ U_{\rm multi}^{[b_\ell]}$ and the numerator of $ U_{\rm multi}^{[a_{\ell}]}$, as well as 
between $ \gamma_{\rm multi}^{[a_\ell]}$ and the numerator of $ U_{\rm multi}^{[b_{\ell+1}]}$, can be observed, as a consequence of sharing the same structure of Legendre polynomials in~\eqref{Umulti} and~\eqref{gammageneral}, respectively.

\begin{center}
\begin{tabular}{ |c||c|c| } 
 \hline
 & $U_{\rm multi}$ & $\gamma_{\rm multi}$ \\ 
 \hline
 \hline
$a_0$ monopole & $\displaystyle{\frac{a_0}{R(r,\theta)}}$ 
& 
$\displaystyle{-\frac{a_0^2 f(r) r^2 \sin ^2\theta }{2 R(r,\theta)^4}}$  \\
\hline
\hline
$b_1$ dipole &  $b_1(r-M)\cos\theta$ & $  -b_1^2 f(r) r^2 \sin ^2\theta$ \\ 
 \hline
 \hline
$a_1$ dipole & $\displaystyle{\frac{a_1(r-M)\cos\theta}{R(r,\theta)^{3}}}$ & $\displaystyle{-\frac{a_1^2 f(r) r^2 \sin ^2\theta[r^2 f(r) (8 \cos^2 \theta -\sin^2\theta)+8 M^2 \cos ^2\theta]  }{4 R(r,\theta)^8}}$ \\ 
  \hline
$b_2$ quadrupole & $\displaystyle{\frac{b_2}{2}  \left[2 M^2 \cos ^2\theta +r^2f(r) \left(2 \cos
   ^2\theta -\sin ^2\theta \right)\right]}$ & $\displaystyle{ -\frac{b_2^2 f(r) r^2 \sin ^2\theta[r^2 f(r) (8 \cos^2 \theta -\sin^2\theta)+8 M^2 \cos ^2\theta]} {4}}$ \\ 
  \hline
\end{tabular}
\end{center}

\section{The Kaluza-Klein oxidization}

The action~\eqref{actionEinsca} is obtained via dimensional reduction of the vacuum Einstein theory in $D=5$,
\begin{equation}
   \mathcal{S}= \frac{1}{16\pi G_5}\int d^{5}\hat{X} \sqrt{-\hat{g}}\hat{R} \ ,
    \label{action5}
\end{equation}
under the Kaluza-Klein ansatz
\begin{equation}
d\hat{s}^2=e^{2{\Phi}/\sqrt{3}}g_{\mu\nu}dx^\mu dx^\nu+ e^{-4{\Phi}/\sqrt{3}}dy^2 \ ,
\end{equation}
where hatted quantities are five dimensional and $\hat{X}^N=(x^\mu,y)$ with $N=(\mu,4)$ and $y$ is the cylindrical coordinate along the fifth dimension. From~\eqref{BHmulti}, we thus obtain the five dimensional Ricci flat solution
\begin{equation}
d\hat{s}^2=e^{2{U_{\rm multi}(r,\theta)}/\sqrt{3}}\left[-f(r)dt^2+e^{2\gamma_{\rm multi}(r,\theta)}\left[\frac{dr^2}{f(r)}+r^2d\theta^2\right]+r^2\sin^2\theta d\varphi^2\right]+ e^{-4U_{\rm multi}(r,\theta)/\sqrt{3}}dy^2 \ .
\end{equation}
Thus, any of the solutions in the previous section can now be mapped into a vacuum $D=5$ solution. For $a_\ell=0=b_\ell$ this is just a black string~\cite{Horowitz:1991cd}. Adding the $D=4$ scalar environment yields a gravitational distortion of the black string in $D=5$. For instance, the scalar counterpart of the Melvin solution~\cite{Cardoso:2024yrb} is mapped into
\begin{equation}
  d\hat{s}^2=e^{2b_1(r-M)\cos\theta/\sqrt{3}}\left[-fdt^2+e^{-b_1^2r^2f \sin^2\theta}\left(\frac{dr^2}{f}+r^2d\theta^2\right)+r^2\sin^2\theta d\varphi^2\right]+ e^{-4b_1(r-M)\cos\theta/\sqrt{3}}dy^2 \ .
  \label{D5uni}
\end{equation}
For $M=0$ 
the Kretschmann scalar is
\begin{equation}
 R_{MNPQ}R^{MNPQ}=   \frac{16}{3} b_1^4 \left(3 b_1^2 r^2 \sin ^2\theta +7\right) e^{\frac{2}{3} b_1 r \left(3 b_1 r \sin ^2\theta-2 \sqrt{3} \cos\theta\right)} \ ,
\end{equation}
and thus it is singular as $r\rightarrow \infty$, as in four dimensions. Likely, all solutions with a Schwarzschild mass $M$ in a scalar environment remain singular, either at the horizon of the black string or at infinity, in $D=5$.

Let us remark, that in $D=5$ one could have started with the ansatz
\begin{equation}
  d\hat{s}^2=e^{2V(\rho,z)}\left[-e^{2U(\rho,z)}dt^2+-e^{-2U(\rho,z)}\left\{e^{2\gamma(\rho,z)}\left(d\rho^2+dz^2\right)+\rho^2 d\varphi^2\right\}\right]+ e^{-4V(\rho,z)}dy^2 \ .
  \label{guess2}
\end{equation}
This $D=5$ geometry has three commuting Killing vector fields. Hence it falls into the general class of  the Weyl problem in $D=5$~\cite{Emparan:2001wk} and, with this ansatz, one expects two harmonic functions. The equations are indeed
\begin{equation}
    \Delta_{\mathbb{E}^3}U(\rho,z)=0 \ , \qquad  \Delta_{\mathbb{E}^3}V(\rho,z)=0 
\end{equation}
on the auxiliary Euclidean 3-space $ds^2_{\mathbb{E}^3}=d\rho^2+\rho^2d\varphi^2+dz^2$ and the two line integrals
\begin{eqnarray}
\gamma_{,\rho}=\rho  \left[(U_{,\rho})^2-(U_{,z})^2+3(V_{,\rho})^2-3(V_{,z})^2\right]
\ , \qquad 
\gamma _{,z}=-6\rho V_{,z} V_{,\rho}-2\rho U_{,z} U_{,\rho} \ ,
\end{eqnarray}
which, with the appropriate rescaling of $V$, coincide with what we obtained in $D=4$, as they should.

\section{Discussion and further generalizations}

The standard Schwarzschild-Melvin solution of Einstein-Maxwell theory 
\begin{equation}
 \mathcal{S}= \frac{1}{16\pi}\int d^{4}x \sqrt{-g}\left(R-\frac{1}{4}F_{\mu\nu}F^{\mu\nu}\right) \ ,
\end{equation}
is described by the following metric and gauge potential $\mathcal{A}$, with $F=d\mathcal{A}$,
\begin{eqnarray}
    ds^2=\Lambda(r,\theta)^2\left(-f(r)dt^2+\frac{dr^2}{f(r)}+r^2d\theta^2\right)+\frac{r^2\sin^2\theta}{\Lambda(r,\theta)^2}d\varphi^2 \ , \qquad 
    \mathcal{A}=\frac{Br^2\sin^2\theta}{2\Lambda(r,\theta)}d\varphi \ ,
\end{eqnarray}
where 
\begin{equation}
\Lambda(r,\theta)=1+\frac{B^2}{4}r^2\sin^2\theta \ .
\end{equation}
The solution is everywhere smooth, albeit with non-standard asymptotics. The Ricci scalar vanishes everywhere, by the conformal symmetry of classical source-free electromagnetism. But, for instance, the Ricci tensor squared reads
\begin{equation}
    R_{\mu\nu}R^{\mu\nu}=\frac{4 B^4  (r-2M \sin^2\theta)^2}{r^2 \Lambda^8} \ ,
\end{equation}
such that for $M=0$ reduces to $R_{\mu\nu}R^{\mu\nu}={4 B^4}/(\Lambda^8)$, being smooth everywhere. A curious feature is that even an empty Melvin Universe can bend light so much as to form a light ring~\cite{Junior:2021dyw}.

The recently reported~\cite{Cardoso:2024yrb} scalar counterpart of the Schwarzschild-Melvin solution of Einstein-scalar theory~\eqref{actionEinsca}, given by eq.~\eqref{solb1}, has some similarity with the Melvin Universe, but a crucial difference, due to infinity being nakedly singular. 

In this paper we have pointed out that the solution in~\cite{Cardoso:2024yrb} is a member of a more general family of Schwarzschild BH/masses inside scalar Universes, and explained their systematic construction, via a mapping to a generalized Weyl problem and using well-known pure gravity solutions. The scalar-vacuum solutions are, however, always singular, either at the BH horizon, or at infinity.\footnote{See~\cite{Astorino:2014mda} for a complementary discussion of stationary axisymmetric solutions in scalar-vacuum.} The singularity of the growing solutions could be dealt with by some appropriate cut-off, even though a clear physical justification would be required. 

Despite the large distance singularity, pure gravity analogues of these type of solutions have physical applications, for instance to study tidal effects on the horizon and horizon distortions~\cite{Annulli:2023ydz}. Additionally, one could explore electrically charged generalizations of these solutions in $D=4$ Kaluza-Klein theory simply by boosting the $D=5$ solutions along the cylindrical coordinate, and rotating generalizations either using the inverse scattering method~\cite{Belinsky:1979mh} or the Newman-Janis algorithm~\cite{Newman:1965tw,Azreg-Ainou:2014pra}.


\begin{acknowledgments}
I am grateful to P. Cunha, J. Nat\'ario, J. Novo and E. Radu for comments on a draft of this paper. This work is supported by the Center for Research and Development in Mathematics and Applications (CIDMA) through the Portuguese Foundation for Science and Technology (FCT -- Fundaç\~ao para a Ci\^encia e a Tecnologia) through projects: UIDB/04106/2020, PTDC/FIS-AST/3041/2020, 2022.04560.PTDC (\url{https://doi.org/10.54499/UIDB/04106/2020}; \url{https://doi.org/10.54499/UIDP/04106/2020}; and \url{http://doi.org/10.54499/PTDC/FIS-AST/3041/2020}; \ and \url{https://doi.org/10.54499/2022.04560.PTDC}). This work has further been supported by the European Horizon Europe staff exchange (SE) programme HORIZON-MSCA-2021-SE-01 Grant No.\ NewFunFiCO-101086251.
\end{acknowledgments}

\bigskip

\bibliographystyle{jhep}
\bibliography{sample}

\end{document}